\documentstyle[12pt]{article}

\def\di{\displaystyle}
\def\&{&\di}
\def\bg{\begin{eqnarray}\begin{array}{rcl}\displaystyle}
\def\eg{\end{array} &\di    &\di   \end{eqnarray}}
\def\bgo{\begin{eqnarray*}\begin{array}{rcl}\displaystyle}
\def\ego{\end{array} &\di    &\di \nonumber  \end{eqnarray*}}

\def\d{{\mbox d}}

\def\T{{\mbox T}}

\def\xv{\mbox{\boldmath$x$}}

\def\yv{\mbox{\boldmath$y$}}

\def\pv{\mbox{\boldmath$p$}}

\def\N{\mbox{l}\!\mbox{N}}

\date{\today}

\def\rene{\renewcommand{\arraystretch}{1.8}}
\rene

 \newcommand{\mysection}[1]{\section{#1}\setcounter{figure}{0}
                    \setcounter{table}{0}\setcounter{equation}{0}}

\begin{document}

\begin{titlepage}

\parindent=12pt
\baselineskip=20pt
\textwidth 15 truecm
\vsize=23 truecm
\hoffset=0.7 truecm

\begin{flushright}
   FSUJ-TPI-15/96 \\
  hep-th/9610048
      \end{flushright}
\par
\vskip .5 truecm
\large \centerline{\bf The Definition of Double Commutators}
\large \centerline{\bf and  Consistency in Free Field Theory} 
\par
\vskip 1 truecm
\normalsize
\begin{center}
{\bf J.~M.~Pawlowski}\footnote{e--mail: jmp@hpxs2.physik.uni-jena.de}\\
\it{Theor.--Phys. Institut, Universit\"at Jena\\ 
Fr\"obelstieg 1\\
D--07743 Jena\\
Germany}
\end{center}
 \par
\vskip 2 truecm
\normalsize
\begin{abstract} 
\noindent Within the framework of generalized functions a general consistent
definition of double commutators is given. This definition respects
the Jacobi identity even if the regularization is removed. The  double
commutator of fermionic currents is calculated in this limit. 
              
\noindent We show that BJL--type prescriptions 
and point--splitting prescriptions for calculating double commutators
fail to give correct results in free field theory.
\end{abstract}
       
\vfill
       
\end{titlepage}

\mysection{Introduction}
It has been known since the days of current algebra that an iterative 
computation of 
double commutators of fermionic currents leads to a violation of the 
Jacobi identity 
in the current algebra \cite{buc,jack}. This violation occurs in anomalous
gauge theories 
(see for example \cite{jack}-\cite{mit}) and also in the quark model, 
where it was observed first \cite{buc}. 
    
\noindent In recent publications the failure of iterative schemes has been
claimed and prescriptions for the calculation of double commutators
were presented which manifestly respect the Jacobi identity
\cite{rot1,leut}. However, it was still an open question, whether
these prescriptions define correctly a double commutator. Since a 
violation of the Jacobi identity even 
takes place in the case of free currents \cite{lev}, it is possible to
study the 
different methods for calculating double commutators in free field theory. 
     
\noindent In this paper we construct double commutators within a general
regularization scheme. This enables us to discuss the consistency of
BJL--type and point--splitting  prescriptions used in the
literature. We claim that there is no BJL--type
prescription, which fulfills the consistency condition 
of a correct regularization. Although point--splitting is
consistent, the averaging procedure normally used leads to
inconsistencies. Since the consistent average procedure is very
complicated the advantages of point splitting are spoiled. 
               
\noindent In the second section we give a definition of double
commutators within a general regularization and calculate them in the
limit, in which the regularization is removed. In the third section 
point--splitting and BJL--type prescriptions are  
discussed. We show that these prescriptions fail to give correct
results in free field theory. We summarize and discuss these results. 
              
\mysection{Double commutators within a general regularization}
In this section we construct a double commutator of free
fermionic currents within a general regularization in four dimensions. 
The explicit form of the free fermionic fields is well-known. We define a 
regularized version as 
\bg
\psi_f(x) &\di   := &\di   \int\d \tilde p\sum_\alpha \Bigl[b_\alpha(p) 
u^\alpha(p)
e^{-ipx}+d^+_\alpha(p) v^\alpha(p)e^{ipx}\Bigr]f(\pv^2)\\\di    
\bar\psi_f(x)     &\di     = &\di     \int\d \tilde p\sum_\alpha
\Bigl[b_\alpha^+
(p)\bar 
u^\alpha(p)e^{ipx}+d_\alpha(p) \bar v^\alpha(p)e^{-ipx}\Bigr]f(\pv^2)
\end{array}& & \label{regulferm}\end{eqnarray}
with
\bgo 
\{b_\alpha^+(p),b_\beta(p')\} &\di   = &\di   (2\pi)^3
2p_0\delta^3(\pv-\pv')
\delta_{\alpha\beta}\\\di 
\{d_\alpha^+(p),d_\beta(p')\} &\di   = &\di   (2\pi)^3
2p_0\delta^3(\pv-\pv')
\delta_{\alpha\beta}\\\di 
\d\tilde p &\di    := &\di    \frac{\d^4 p}{(2\pi)^3} \delta(p^2)\theta(p_0).
\ego   
In eq. (\ref{regulferm}) $f$ is an arbitrary function of fast decrease
with 
\bg
f(0) &\di   = &\di   1\\\di 
\lim_{\|\pv\|\rightarrow \infty}\|\pv\|^n f(\pv^2) &\di   = &\di   0\
\ \ \forall n
\in \N.
\label{general}\eg 
The regularized free vector and axial currents are given by 
\bg
V^\mu(x;f) &\di     := &\di     :\bar\psi_f(x)\gamma^\mu\psi_f(x):\\\di   
A^\mu(x;f) &\di     := &\di     :\bar\psi_f(x)\gamma_5\gamma^\mu\psi_f(x):
\end{array}&\di   &\di   \label{regul:01}\end{eqnarray} 
We define the equal--time double commutators  $\Gamma_l^{ij},\ l\ =\
1,2,3$ of these currents as the
regularization limit $f\rightarrow 1$ of 
\bg
\Gamma_1^{ij}(\xv_1,\xv_2,\xv_3;f) &\di     := &\di     
\langle
0|\Bigl[A^0(x_1;f),[V^i(x_2;f),V^j(x_3;f)]\Bigr]_{ET}|0\rangle\\\di  
\Gamma_2^{ij}(\xv_1,\xv_2,\xv_3;f) &\di     := &\di
\langle
0|\Bigl[V^j(x_3;f)
,[A^0(x_1;f),V^i(x_2;f)]\Bigr]_{ET}|0\rangle\\\di   
\Gamma_3^{ij}(\xv_1,\xv_2,\xv_3;f) &\di     := &\di
\langle
0|\Bigl[V^i(x_2;f)
,[V^j(x_3;f),A^0(x_1;f)]\Bigr]_{ET}|0\rangle.
\end{array}&\di    &\di    \label{dop:11}\end{eqnarray}
The Jacobi identity is fulfilled by these regularized 
double commutators. 
\bg
J &\di   := &\di
\Gamma_1^{ij}(\xv_1,\xv_2,\xv_3;f)+\Gamma_2^{ij}(\xv_1,\xv_2,\xv_3;
f)+\Gamma_3^{ij}(\xv_1,\xv_2,\xv_3;f)\ =\ 0.
\eg
We have 
\bg
\Gamma_l^{ij}(\xv_1,\xv_2,\xv_3) \& := \& \lim_{f\rightarrow
  1}\Gamma_l^{ij}(\xv_1,\xv_2,\xv_3;f).
\eg
The equal--time double commutators are sums of products of 
the spatial $\delta$--distribution and its derivatives. The structure
of the 
$\Gamma_l^{ij},\ l\ =\ 1,2,3$ is fixed by the symmetry properties of the 
corresponding double commutator. For example, $\Gamma^{ij}_3$ is given by
\bg
\Gamma^{ij}_3(\xv_1,\xv_2,\xv_3) &\di    = &\di    2i\epsilon^{ijk}\Bigl[
a_1\partial_k\delta(\xv_1-\xv_2)\delta(\xv_3-\xv_2)\\\di  
       &\di      &\di
       +a_2\Delta\partial_k\delta(\xv_1-\xv_2)\delta(\xv_3-\xv_2)
\\\di 
&  &\di +a_3\partial_k\delta(\xv_1-\xv_2)
\Delta\delta(\xv_3-\xv_2)\\\di 
& &\di
+a_4\partial_k\partial_l\delta(\xv_1-\xv_2)\partial_l\delta(\xv_3-\xv_2)
\Bigr]\\\di  
&\di     &\di    -((\xv_1-\xv_2)\leftrightarrow (\xv_3-\xv_2)).
\end{array}\label{111:dk}\end{eqnarray}
The factor $a_1$ is dependent on the choice of the regularization
function $f$ because it has not zero dimension \cite{rot1,janD}. This
is already known from similar coefficients in commutators of free
currents. We
determine the other components by calculating the following integrals
($k$ is fixed). 
\bg
 \langle\xv_1^2 (\xv_1)_k\rangle  
&\di    :=  &\di    \int\d^3 x_1\d^3 x_3 \ \xv_1^2(\xv_1)_k\bar
\Gamma_k^3[\xv_1,\xv_3;f]\\\di  
  \langle\xv_1^2 (\xv_3)_k\rangle
&\di    := &\di     \int\d^3 x_1\d^3 x_3 \ \xv_1^2(\xv_3)_k\bar 
\Gamma_k^3[x_1,0,x_3;f]\\\di  
  \langle(\xv_1)_k(\xv_1)_l (\xv_3)_l\rangle  &\di    := &\di
  \int\d^3 x_1\d^3 x_3 \ (\xv_1)_k(\xv_1)_l(\xv_3)_l\bar
\Gamma_k^3[x_1,0,x_3;f]
\end{array}\label{gleich:10}\end{eqnarray}
with 
\bg
2i\epsilon^{ijk}\bar\Gamma_k^3[\xv_1,\xv_2,\xv_3;f] &\di   := &\di
\Gamma_3^{ij}
[\xv_1,\xv_2,\xv_3;f] .  
\eg 
These integrals are connected with the coefficients $a_j,\ j\neq 1$ by 
\bg
\langle\xv_1^2 (\xv_1)_k\rangle &\di    = &\di    -10 a_2\\\di  
2\langle\xv_1^2 (\xv_3)_k\rangle- \langle(\xv_1)_k(\xv_1)_l(\xv_3)_l\rangle  
&\di = &\di    
-10 a_3\\\di  
3\langle(\xv_1)_k(\xv_1)_l(\xv_3)_l\rangle-\langle\xv_1^2(\xv_3)_k\rangle  
&\di    = &\di    
-10 a_4.
\end{array}\label{gleich:11}\end{eqnarray}
The $f$-dependence of the integrals in eq. (\ref{gleich:10}) is of the type 
\bg
\int\d^3 p\frac{\pv_i \pv_j}{(\pv^2)^{\frac{3}{2}}}
\frac{\partial}{\partial \pv^2}f(\pv^2) &\di = &\di
\frac{2\pi}{3}\delta_{ij}f(0)\ = \ \frac{2\pi}{3}\delta_{ij}.
\label{structure}\eg
Thus only $a_1$ depends on $f$ and it follows
 \bg
  \Gamma^{ij}_3(\xv_1,\xv_2,\xv_3) &\di    = &\di
  -\frac{1}{81\pi^2}
\epsilon^{ijk}
\Bigl\{c[f]\partial_k
\delta(\xv_1-\xv_3)[\delta(\xv_1-\xv_2)+\delta(\xv_3-\xv_2)]\\\di  
                     &\di       &\di    \ \
                     +\frac{22}{5}\partial_k\Delta
\delta(\xv_1-\xv_3)
[\delta(\xv_1-\xv_2)+\delta(\xv_3-\xv_2)]\\\di  
                     &\di       &\di    \  \
                     +2\Bigl(\Delta\delta(\xv_1-\xv_2)
\partial_k
\delta(\xv_3-\xv_2)\\\di 
& &\di -\Delta\delta(\xv_3-\xv_2)\partial_k\delta(\xv_1-\xv_2)
\Bigr)\\\di  
                     &\di       &\di    \  \
                     -2\Bigl(\partial_k\partial_l
\delta(\xv_1-\xv_2)
\partial_l\delta(\xv_3-\xv_2)\\\di 
&\di    &\di   -\partial_k\partial_l\delta(\xv_3-\xv_2)
\partial_l\delta(\xv_1-\xv_2)\Bigr)\Bigr\}, 
\label{result1}\eg
where $c[f]$ is an $f$--dependent constant  ($c[f]= -162 \pi^2 i a_1[f]$). 
$\Gamma^{ij}_2$ and $\Gamma^{ij}_1$ follow similarly 
\bg
 \Gamma^{ij}_2(\xv_1,\xv_2,\xv_3)  &\di    = &\di
 -\frac{1}{81\pi^2}
\epsilon^{ijk}
\Bigl\{c[f]\partial_k
\delta(\xv_1-\xv_2)[\delta(\xv_1-\xv_3)+\delta(\xv_2-\xv_3)]\\\di  
                     &\di       &\di    \ \
                     +\frac{22}{5}\partial_k\Delta
\delta(\xv_1-\xv_2)
[\delta(\xv_1-\xv_3)+\delta(\xv_2-\xv_3)]\\\di  
                     &\di       &\di    \  \
                     +2\Bigl(\Delta\delta(\xv_1-\xv_3)
\partial_k
\delta(\xv_2-\xv_3)\\\di 
&  &\di -\Delta\delta(\xv_2-\xv_3)\partial_k\delta(\xv_1-\xv_3)
\Bigr)
\\\di                      &\di       &\di    \  \
-2\Bigl(\partial_k\partial_l
\delta(\xv_1-\xv_3)
\partial_l\delta(\xv_2-\xv_3)\\\di 
&\di    &\di   -\partial_k\partial_l\delta(\xv_2-\xv_3)
\partial_l\delta(\xv_1-\xv_3)\Bigr)\Bigr\}
\label{result2}\eg
and 
\bg
\Gamma^{ij}_1(\xv_1,\xv_2,\xv_3) &\di    = &\di
-\Bigl[\Gamma^{ij}_2(\xv_1,\xv_2
,\xv_3)+
\Gamma^{ij}_3(\xv_1,\xv_2,\xv_3)\Bigr].
\end{array}\label{111:jaco}\end{eqnarray}
The double commutators given in
the literature \cite{rot1,leut} are not compatible with these results.
Neither the use of BJL--type
prescriptions nor the use of point--splitting prescriptions lead to
the results (\ref{result1},\ref{result2},\ref{111:jaco}).  
\section{Comparison with point--splitting and BJL--type\\ prescriptions}
At the beginning we introduce an extension of the regularization 
given in eq. (\ref{dop:11}), where
some of the properties of a double commutator are lost. We define a 
more general expression $\Gamma_l^{ij}$ as follows. 
For convenience we choose as an explicit regularization 
\bg
f_\epsilon(\pv^2) &\di   := &\di   \exp\{-\epsilon \|\pv\|\}.
\end{array}& & \label{epsilon}\end{eqnarray}
In order to compare the result with BJL--type prescriptions 
we also introduce a regularization for the equal--time limits.
\bg
\Gamma_1^{ij}(\xv_1,\xv_2,\xv_3;\epsilon,r) &\di     := &\di     
\int\prod_{i=1}^3\d x_i^0\langle
0|\Bigl[A^0(x_1;f_\epsilon),[V^i(x_2;f_\epsilon),V^j(x_3;f_\epsilon)]\Bigr]
|0\rangle\delta_r(x^0,t)\\\di  
\Gamma_2^{ij}(\xv_1,\xv_2,\xv_3;\epsilon,r) &\di     := &\di
\int\prod_{i=1}^3\d x_i^0\langle
0|\Bigl[V^j(x_3;f_\epsilon)
,[A^0(x_1;f_\epsilon),V^i(x_2;f_\epsilon)]\Bigr]|0\rangle
\delta_r(x^0,t)\\\di   
\Gamma_3^{ij}(\xv_1,\xv_2,\xv_3;\epsilon,r) &\di     := &\di
\int\prod_{i=1}^3\d x_i^0\langle
0|\Bigl[V^i(x_2;f_\epsilon)
,[V^j(x_3;f_\epsilon),A^0(x_1;f_\epsilon)]\Bigr]|0\rangle
\delta_r({x^0},t)\\\di  
\delta_r({x^0},t) &\di    := &\di    \prod_{i=1}^{3}\delta_{r_i}(x_i^0-t).
\end{array}&\di    &\di    \label{dop:12}\end{eqnarray}
The $\delta_{r_i}$ are regularizations of the one--dimensional 
$\delta$--distribution with regularization parameters $r_i$. 
The Jacobi identity is fulfilled by these expressions
\bg
J &\di   := &\di
\Gamma_1^{ij}(\xv_1,\xv_2,\xv_3;\epsilon,r)+\Gamma_2^{ij}(\xv_1,\xv_2,\xv_3;
\epsilon,r)+\Gamma_3^{ij}(\xv_1,\xv_2,\xv_3;\epsilon,r)\ =\ 0.
\eg
We want to emphasize that we have introduced independent
regularizations $r_i$  of the
equal--time limits only to compare our results with other methods used in the
literature (e.g. BJL--limit, see appendix). The symmetry properties of 
the double commutators (under the interchange of currents) force us to take 
\bg
r_1\ = \ r_2 &\di    = &\di     r_3\ =\ \tilde r.
\end{array}&\di    &\di   \label{equal}\end{eqnarray}
Furthermore, it is quite natural to calculate equal--time double commutators
of the 
regularized currents (keeping $f$ fixed and remove $r$ first) and to take
the regularization limit of $f\rightarrow 1$ after this computation. 
This corresponds to the usual definition of quantities in a
regularized theory. 
                  
\noindent First we concentrate on the case of eq. (\ref{equal}), which
still leads to an expression consistent with the symmetry properties
of an double commutator. For the
calculation we need an explicit expression for the regularized 
$\delta$--distribution, 
\bg
\delta_{\tilde r}(x_i^0) &\di   =  &\di
\frac{1}{\pi}\frac{\tilde r}{\tilde r^2+{x_i^0}^2}\frac{a^4}{(a^2
+{x_i^0}^2)^2}.
\end{array}\label{def:01}\end{eqnarray}
Here $a$ is a free parameter with $a\ \neq \ 0$. The limit, in which
the regularization is removed does not depend on $a$. 
It is only introduced for technical reasons. Performing the integration of
$\Gamma_3^{ij}(\xv_1,\xv_2,\xv_3;\epsilon,r)$ as in eq.
(\ref{gleich:10}), the coefficients $a_j,\ j\ =\ 2,3,4$ follow with 
(\ref{gleich:11}) after a long but straightforward calculation \cite{janD}. 
\bg
a_2(u) &\di    = &\di    \frac{i}{30\pi^2}\Biggl(\frac{2+3u}{3+4u}+
\frac{4+18u+25u^2+10u^3}{(3+4u)^3}\Biggr)\\\di  
a_3(u) &\di    = &\di    -\frac{i}{30\pi^2}\Biggl(\frac{u}{3+4u}+
\frac{10+40u+49u^2+19u^3}{(3+4u)^3}\Biggr)\\\di  
a_4(u) &\di    = &\di    \frac{i}{30\pi^2}\Biggl(-2
\frac{u}{3+4u}+\frac{10+35u+32u^2+2u^3}{(3+4u)^3}\Biggr)
\label{coefficients}\eg
with 
\bg
u & = &\di \frac{\tilde r}{\epsilon}.
\eg 
The coefficients $a_j$ depend on $u$, which is a measure for the
mixing of the 
regularization limit ($\epsilon\rightarrow 0$) with the equal--time
limit ($\tilde r\rightarrow 0$). The convenient choice $u\ =\ 0$ (first 
taking the equal-time limit and then removing the regularization of the 
currents) has been calculated already in the previous section. However, 
even for arbitrary $u$ the results in the literature  \cite{rot1,leut} are 
not compatible with eq. (\ref{coefficients}). 
                     
\noindent Now we discuss how point-splitting and BJL--prescriptions fit into 
this picture. We point out that the limit procedure in a 
point--splitting prescription should correspond to the case $u\ =\ 0$. 
Taking the value of $u$ equal to zero is equivalent to a fixed--time
regularization. The use
of regulator functions $f$ with dependence only on spatial momenta is
an introduction of 
point--splitting as an spatial regularization in an unambiguous way. 
The averaging, relevant in point--splitting 
\bg
\frac{\epsilon_i\epsilon_j}{\epsilon^2} &\di  = &\di  \delta_{ij}\ \
\mbox{for 
commutators}\\\di 
\frac{\epsilon_i\epsilon_j\epsilon_k\epsilon_m}{\epsilon^4} &\di  = &\di  
\frac{1}{3}(\delta_{ij}\delta_{km}+\delta_{ik}\delta_{jm}+\delta_{im}
\delta_{jk})\ \ \mbox{for double commutators}
\eg
is replaced by the explicit calculation of these quotients (see
(\ref{structure})). 
The point--splitting result for the commutator is reproduced in our
scheme \cite{janD}. For the double commutator the provided averaging leads 
to a different result as the explicit calculation of the quotient in the
general regularization (eq. (\ref{dop:11})). We conclude that
the usual averaging procedure is not applicable to calculate 
double commutators. 
       
\noindent In the following we concentrate on BJL--type prescriptions 
for double commutators. To reproduce the results of the BJL--type 
prescriptions, we have to use the extension (eq. (\ref{dop:12})) of the 
general regularization (eq. (\ref{dop:11})). The calculations involved are 
tedious and not illuminating, so we only discuss the results. 
First we reproduce the iterative BJL--scheme 
within eq. (\ref{dop:12}), for which we need a general regularization of 
the equal-time limits with different $r_i$
(see eq. (\ref{dop:12})).  The result of the iterative BJL--scheme 
follows by the following different limits
\cite{janD} (see appendix for the definition of the BJL--limit) 
\bg
(q_0,k_0)[A^0(x_1),V^i(x_2),V^j(x_3)] &\di    = &\di     
(r_{1},r_{2},r_{3},\epsilon)[\Gamma^{ij}_1(x_1,x_2,x_3;\epsilon,r)]\\\di  
\mbox{formally}\rightarrow\ \ \ &\di {=} &\di \langle
0|\Bigl[A^0(x_1),[V^i(x_2),V^j(x_3)]
\Bigr]_{ET}|0\rangle\\\di
(k_0,p_0)[A^0(x_1),V^i(x_2),V^j(x_3)] &\di    = &\di    
(r_{3},r_{1},r_{2},\epsilon)[\Gamma^{ij}_2(x_1,x_2,x_3;\epsilon,r)]\\\di 
\mbox{formally}\rightarrow\ \ \ &\di {=} &\di     \langle
0|\Bigl[V^j(x_3),[A^0(x_1),V^i(x_2)]\Bigr]_{ET}|0\rangle\\\di   
-(q_0,p_0)[A^0(x_1),V^i(x_2),V^j(x_3)] &\di    = &\di    
 (r_{2},r_{3},r_{1},\epsilon)[\Gamma^{ij}_3(x_1,x_2,x_3;\epsilon,r)]\\\di 
\mbox{formally}\rightarrow\ \ \ &\di {=} &\di    \langle
0|\Bigl[V^i(x_2),[V^j(x_3),
A^0(x_1)]\Bigr]_{ET}|0\rangle  
\end{array}\label{aqui:01}\end{eqnarray}
with
\bgo
(r_{i_1},r_{i_2},r_{i_3},\epsilon)[\Gamma_l^{ij}] \& := \& 
\lim_{r_{i_1}\rightarrow 0}\lim_{r_{i_2}\rightarrow 0}\lim_{r_{i_3}
\rightarrow 0}
\lim_{\epsilon\rightarrow 0}[\Gamma_l^{ij}].
\ego
The reason for this equality between the iterative BJL--limits and the 
$(r_{i_1},r_{i_2},r_{i_3},\epsilon)$-limits is the following. 
Clearly, the regularization parameter $\epsilon$, which 
corresponds to a regularization of the contributing $n$--point functions is 
not involved explicitly in the iterative scheme (in every space--time
prescription, e.g. BJL--type prescriptions) 
so we set it to zero first\footnote{In fact, implicitly $\epsilon$ is
  involved; but it is only used to get formally definiteness of the taken
  equal--time limits.}. Then we have to perform the limits $r_i$. 
Each limit $r_i$ corresponds to the limit $(x_i)_0\rightarrow t$ which
is equivalent to the limit $q_i^0\rightarrow\infty$ with $q_i$ is the
corresponding momenta of $x_i$. Thus the limits on the right side of
(\ref{aqui:01}) reproduce the
iterative BJL--type limits. However, we are not
allowed to calculate each double commutators $\Gamma_l^{ij}$ with a
different limit, if the results depend on the chosen limit. This
is the reason for the violation of the Jacobi identity. Furthermore
the $\Gamma_l^{ij}$ with $r_1\ \neq r_2\ 
\neq r_3$ loose the symmetry properties a double commutator need to
have under the interchange of the currents. We take as an example
$\Gamma_3^{ij}$ calculated in the iterative scheme 
\bg
{\Gamma_3^{ij}}_{iterat}(\xv_1,\xv_2,\xv_3) \&:= \& (r_{2},r_{3},r_{1},
\epsilon)
[\Gamma^{ij}_3(\xv_1,\xv_2,\xv_3;\epsilon,r)].
\label{iterativ1} \eg
We get the following inequality 
\bg
{\Gamma_3^{ji}}_{iterat}(\xv_1,\xv_3,\xv_2) \& \neq \& (r_{2},r_{3},
r_{1},\epsilon)
[\Gamma^{ji}_3(\xv_1,\xv_3,\xv_2;\epsilon,r)],
\label{inconsistent}\eg
which should be an equality for a consistent scheme. 
However, whereas the left-hand side is determined by eq.
(\ref{iterativ1}) with finite coefficients $a_i,\ i=2,3,4$, 
the coefficients $a_i'$ at the right-hand side of eq.
(\ref{inconsistent}) are not only different
but also diverging \cite{janD}. If one insists nevertheless on using one of
the limits $(r_{i_1},r_{i_2},r_{i_3},\epsilon)$ in
(\ref{aqui:01}) for the calculation of the double commutators, the only
prescription, which respects the Jacobi identity
is to calculate all $\Gamma_l^{ij}$ within this limit. However, two of
the $\Gamma_l^{ij}$ are not well defined in this procedure. The
coefficients $a_i,\ i =2,3,4$ diverge. This also shows the invalidity of 
the iterative scheme. Only for $r_i=r,\ i=1,2,3$ the symmetry properties 
are maintained. 
         
With these remarks it is possible to look at general BJL--type
prescriptions to calculate double commutators $[A(x),[B(y),C(z)]]$.
All these prescriptions have to use different
equal--time limits $(r_{i_1},r_{i_2},r_{i_3},\epsilon)$ in calculating 
the different double commutators (see as
an example the iterative scheme (\ref{aqui:01}), appendix). If the
Jacobi identity is 
violated by the iterative scheme this indicates that different 
limiting procedures do not lead to an unique result. As we have
explained in this case every single result is incompatible with the
symmetry properties of a double commutator. It follows that
in this case all BJL--type scheme are using inconsistent limit procedures. 
Even a combination of limits that leads to expressions 
which respect the Jacobi identity provides not a 
consistent scheme in general because each limit itself is inconsistent. 
     
\mysection{Conclusion}

We have calculated double commutators of free vector and axial currents
within a general regularization (see eq. (\ref{dop:11})). The Jacobi identity 
is fulfilled within this framework and the dimensionless coefficients $a_i,\
i=2,3,4$ are independent of the chosen limit, in which the
regularization is removed ($f\rightarrow 1$). Only
$a_1[f]$ depends on the regularization, which is well known from the
calculation of commutators of free currents.  
                         
\noindent Neither point--splitting results nor the results of BJL--type
prescriptions are compatible with the double commutators (see eq.
(\ref{result1},\ref{result2},\ref{111:jaco})). 
                
\noindent In the general regularization (eq. (\ref{dop:11})) the
averaging procedure relevant in
point--splitting is replaced by an explicit calculation of the
quotients 
$$
\frac{\epsilon_i\epsilon_j}{\epsilon^2} \ , \ 
\frac{\epsilon_i\epsilon_j\epsilon_k\epsilon_m}{\epsilon^4}. 
$$
Thus we conclude, that the usual averaging is inconsistent in the case
of double commutators. However the advantage of point--splitting is
the simplicity of the involved calculations due to the averaging
procedure. This advantage is spoiled, if one
has to calculated the quotients explicitly.  
                         
\noindent To reproduce the results of BJL--type prescriptions 
used in in the literature, we have to introduce an extension of the
general consistent regularization (eq. \ref{dop:12}), for which 
consistency conditions for double commutators are violated. We
have shown explicitly, how to reproduce the iterative BJL--prescription
within this extension. BJL--type prescriptions have to use
different equal--time limits for the calculation of different double
commutators. However, if the result depends on the limit, this is
inconsistent. The violation of the Jacobi identity
in the iterative scheme is the result of the use of different 
equal--time limits for each double commutator. This violation measures 
the consistency of the used methods and therefore it has no physical
interpretation. Furthermore it shows the general inconsistency of BJL--type
methods for calculating double commutators. It is due
to the fact that in every calculation of double commutators with
BJL--type prescription one is forced to take different equal--time
limits for each double commutator. 
              
We conclude that the use of point--splitting--prescrip\-tions 
related with the iterative scheme and BJL--type methods leads 
to wrong results if the iterative schemes violate the Jacobi
identity. This is true also 
in interacting theories like chiral gauge theories. However, in these
theories it is possible to use their geometric structure for an
algebraic BJL--aproach to 
compute double commutators with anomalous Schwinger terms \cite{janD}.

\begin{appendix}
\section{Appendix}     
The Bjorken--Johnson--Low (BJL) limit is a prescription often used to
calculate commutators \cite{buc,jack,bjork}. It does not suffer from
averaging procedures like the fixed--time prescriptions like the
point--splitting. We define a BJL--type 
prescription to calculate double commutators in the following way.
\bg
 \langle 0| \Bigl[A(x),[B(y),C(0)]\Bigr]_{ET}|0\rangle &\di   = &\di
 \lim_{p_0
\rightarrow \infty}
\lim_{q_0\rightarrow \infty} p_0 q_0 T(\xv,\yv;p_0,q_0)\\\di  
T(\xv,\yv;p_0,q_0) &\di   := &\di   \int\d x_0 \d y_0 e^{ip_0 x_0}e^{iq_0 y_0}
\langle 0|\T A(x)B(y)C(0)|0\rangle
\eg
We can easily see that this definition provides not only one
prescription to 
calculate double commutators \cite{rot1}. We define the following different 
limits  
\bg 
(g,l)[A(x),B(y),C(0)] &\di   := &\di   \lim_{g\rightarrow
  \infty}\lim_{l
\rightarrow \infty}gl T(\xv,\yv;p_0,q_0)\\\di  
g,l &\di   \in &\di   \{p_0,q_0,-(p_0+q_0)\}.
\eg
\end{appendix}

\end{document}